\documentclass[12pt]{article}
\usepackage{epsf}
\usepackage{a4}
\usepackage{amsfonts}		% open math symbols
\usepackage{cite}		% collapse citations
\usepackage{color}
\usepackage{multirow}

\def\hybrid{\topmargin 0pt
        \oddsidemargin 0pt
        \headheight 0pt \headsep 0pt
        \textwidth 6.25in       % A4 paper
        \textheight 9.5in       % A4 paper
        \marginparwidth .875in
        \parskip 5pt plus 1pt   \jot = 1.5ex}

\catcode`\@=11
\def\marginnote#1{}

\newcount\hour
\newcount\minute
\newtoks\amorpm
\hour=\time\divide\hour by60
\minute=\time{\multiply\hour by60 \global\advance\minute by-\hour}
\edef\standardtime{{\ifnum\hour<12 \global\amorpm={am}%
        \else\global\amorpm={pm}\advance\hour by-12 \fi
        \ifnum\hour=0 \hour=12 \fi
        \number\hour:\ifnum\minute<10 0\fi\number\minute\the\amorpm}}
\edef\militarytime{\number\hour:\ifnum\minute<10 0\fi\number\minute}

\def\draftlabel#1{{\@bsphack\if@filesw {\let\thepage\relax
   \xdef\@gtempa{\write\@auxout{\string
      \newlabel{#1}{{\@currentlabel}{\thepage}}}}}\@gtempa
   \if@nobreak \ifvmode\nobreak\fi\fi\fi\@esphack}
        \gdef\@eqnlabel{#1}}
\def\@eqnlabel{}
\def\@vacuum{}
\def\draftmarginnote#1{\marginpar{\raggedright\scriptsize\tt#1}}

\def\draft{\oddsidemargin -.5truein
        \def\@oddfoot{\sl preliminary draft \hfil
        \rm\thepage\hfil\sl\today\quad\militarytime}
        \let\@evenfoot\@oddfoot \overfullrule 3pt
        \let\label=\draftlabel
        \let\marginnote=\draftmarginnote
   \def\@eqnnum{(\theequation)\rlap{\kern\marginparsep\tt\@eqnlabel}%
\global\let\@eqnlabel\@vacuum}  }

\def\numberbysection{\@addtoreset{equation}{section}
        \def\theequation{\thesection.\arabic{equation}}}

\def\titlepage{\@restonecolfalse\if@twocolumn\@restonecoltrue\onecolumn
     \else \newpage \fi \thispagestyle{empty}\c@page\z@
        \def\thefootnote{\fnsymbol{footnote}}
	\setcounter{page}{0} }
\def\endtitlepage{\if@restonecol\twocolumn \else  \fi
        \def\thefootnote{\arabic{footnote}}
        \setcounter{footnote}{0}}  %\c@footnote\z@ }
\definecolor{c1}{rgb}{1, 0, 0}
\definecolor{c2}{rgb}{0, 1, 0}
\definecolor{c3}{rgb}{0, 0, 1}
\definecolor{c4}{rgb}{1, 0, 1}
\definecolor{c5}{rgb}{0, 1, 1}

\catcode`@=12
\relax

\def\ie{\hbox{\it i.e.}}
\def\nn{\nonumber}
\def\beq{\begin{equation}}
\def\eeq{\end{equation}}
\def\bea{\begin{eqnarray}}
\def\eea{\end{eqnarray}}
\relax
\numberbysection
\hybrid
\begin{document}
\begin{titlepage}
\begin{center}
{\large\bf
Geometrical properties of parafermionic spin models.
%Fractal dimensions for parafermions.
}\\[.3in] 

{\bf M.\ Picco$^{1}$, R.\ Santachiara$^{2}$ and A.\ Sicilia$^{1}$}\\
        % {\bf (1)}
	$^1$ {\it LPTHE\/}\footnote[1]{Unit\'e mixte de recherche du CNRS 
UMR 7589.}, % \\
        {\it  Universit\'e Pierre et Marie Curie-Paris6\\
              Bo\^{\i}te 126, Tour 24-25, 5 \`eme \'etage, \\
              4 place Jussieu,
              F-75252 Paris CEDEX 05, France, \\
    e-mail: {\tt picco,sicilia@lpthe.jussieu.fr}. }\\
        %{\bf (2)}
	$^2$ {\it LPTMS, \footnote[3]{Unit\'e mixte de 
             recherche du CNRS UMR 8626.}, % \\
             Universit\'e Paris-Sud\\
             B\^atiment 100\\
             91405 Orsay, France. \\
    e-mail: {\tt santachi@lpt.ens.fr}. }\\
\end{center}
%\vskip .04in
\centerline{(Dated: \today)}
\vskip .2in
\centerline{\bf ABSTRACT}
\begin{quotation}

  We present measurements of the fractal dimensions associated to the
  spin clusters for $Z_4$ and $Z_5$ spin models. We also
  attempted to measure similar fractal dimensions for the generalised
  Fortuin Kastelyn (FK) clusters in these models but we discovered
  that these clusters do not percolate at the critical point of the
  model under consideration. These results clearly mark a difference in the
  behaviour of these non local objects compared to the Ising model or
  the $3$-state Potts model which corresponds to the simplest cases of
  $Z_N$ spin models with $N=2$ and $N=3$ respectively. We compare
  these fractal dimensions with the ones obtained for SLE interfaces.
\vskip 0.5cm
\noindent
%\pacs
%{PACS numbers: 75.50.Lk, 05.50.+q, 64.60.Fr}

% PACS 1999 classification: http://www.aip.org/pacs/pacs99/pacscheme.html

\end{quotation}
\end{titlepage}

\section{Introduction}
The geometrical description of phase transitions has a long history
\cite{Duplantier}. The existence of exactly solved models and, most
importantly, the richness of the conformal symmetry in two dimensions
(2D), make the two-dimensional statistical systems an ideal framework
to study this problem. The critical points of 2D systems can be
classified using conformal field theories (CFTs) which also provide a
powerful approach to compute exactly correlation functions of local
operators.  The first major breakthrough in the study of conformally
invariant interfaces in 2D critical models has come from the
introduction of the so-called Coulomb-gas (CG) formalism
\cite{Nienhuis_CG}. When a model is provided with a CG description,
the combination of CG and CFT techniques allow the exact computation
of geometrical exponents which characterize the fractal shape of
critical interfaces. This has been done for a variety of critical
statistical models as critical percolation, self-avoiding walks, loop
erased random walks, etc. All these models are associated to the so
called minimal CFTs or equivalently to the critical phases of $O(n)$
models \cite{Nienhuis_CG}.  Using the CG description of the $O(n)$
model, the fractal dimension (and more generally all the multi-fractal
scaling exponents) of critical interfaces has been exactly computed
\cite{Saleur,Duplantier2}.  A remarkable recent development in the
study of critical interfaces in 2D systems came with the introduction
of a conceptually new approach based on the so called Schramm-Loewner
evolutions (SLEs), which are growth processes defined via stochastic
evolution of conformal maps
\cite{Walter,Cardy_review,Bernard_review}. Again, the SLE approach,
which provides a geometrical description of CFT, is completely
understood only in the case of minimal CFTs
\cite{Bernard_connection1,Bernard_connection2,Bernard_connection3}.

The minimal CFTs are constructed by demanding the correlation
functions to satisfy the conformal symmetry alone and they represent a
small set of CFT theories.  There are many other interesting CFTs,
called extended CFTs, which describe many condensed matter and
statistical problems which are characterized in general by some
internal symmetry such as, e.g., the $SU(2)$ spin-rotational symmetry
in spin chains \cite{Affleck} or replica permutational symmetry in
disordered systems \cite{Ludwig,DPP}.  Despite all the recent activity
and progress, the geometrical properties of such extended CFTs are in
general not understood.  In this respect some progress has been done
by studying the connection between SLE and Wess-Zumino-Witten models,
i.e. CFTs with additional Lie-group symmetries
\cite{Rasmussen,Ludwig2}, and by defining loop models associated to
some extended CFTs \cite{Pasquier,DFSZ,Fendley2006,Cardy2008,
Rajapour}. In this direction of investigation, a very interesting
family of critical models are the $Z_N$ spin models (defined below)
i.e. a lattice of spins which can take $N$-values. The
nearest-neighbor interaction defining the model is invariant under a
$Z_N$ cyclic permutation of the states. For $N=2$ and $N=3$ these
models correspond to the well known Ising and three-states Potts model
whose critical points are described by minimal CFTs. For $N\geq 4$
instead, the models admit critical points described by parafermionic
theories which are extended CFTs where the role of the $Z(N)$ symmetry
beside the one of conformal symmetry must be taken into account.  The
geometrical properties of $Z_N$ spin models are for many aspects
unknown and their study are expected to provide general deep insights
on the geometrical description of extended CFTs.
 
In a recent work \cite{Raoul} one of the authors proposed an extension
of the concept of SLE to the case of the $Z_N$ parafermionic theory.
An SLE interface is associated to the (conformal) boundary condition
which generates it.  Considering the $Z_N$ spin model on a bounded
domain, say the half-plane for instance, and specifying a particular
boundary condition, an SLE interface was identified as the boundary
of the spin cluster connected to the negative axis
\cite{Raoul}. By the term spin cluster we mean the cluster
which connects spins with equal value. This interface was further
studied in \cite{MarcoRaoul} where the corresponding fractal dimension
has been shown to be in agreement with the CFT predictions in
\cite{Raoul}. Nevertheless by considering other boundary conditions,
we obtained different results for the fractal dimensions of the
corresponding interfaces in the case of the $Z_N$ spin model with
$N\geq 4$ \cite{MarcoRaoul2}.  This is at odd from results for the
Potts models for which a single fractal dimension for the spin
interface (i.e. the boundary of the spin cluster) is obtained
\cite{Gamsa}.
 
The present work is thus motivated by determining the bulk fractal
dimension, \ie the fractal dimension of the boundaries of finite clusters in
the model. To be more precise, we will consider in this work the
fractal dimensions obtained by constructing the distribution of all
the finite closed spin clusters. As we will show later, the
spin clusters do percolate at the critical point of the $Z_N$ spin
model in the sense that there is a one large cluster which span the
entire lattice at the critical point. The distribution of the smallest
clusters can then be used to define a set of exponents as is the case
in percolation theory and from these exponents one can determine the
fractal dimension.

At this point, one needs to provide some explanation on why we
concentrate on the spin clusters.  It is well known that while
the spin clusters do percolate at the critical point in two
dimensional Potts models, they do not contain the physical information
of the model considered.  For example the exponents obtained from the
spin clusters of the two dimensional Ising model are not the
exponents of the Ising model \cite{Coniglio,Sykes}. These exponents
are in fact encoded in some other objects, the FK clusters.
Obviously, the FK clusters must also percolate at the critical
point. That the spin clusters percolate for the same critical
temperature as for the FK clusters is true for the Potts models in two
dimensions but is not a general result. In three dimensions for the
Ising model, the percolation of the Ising model occurs at a different
temperature \cite{Muller,DHMMPW}. In fact, one has no reason to expect
that the spin clusters do percolate at the critical point for
any two dimensional model. That it is the case for the Potts models
can be traced back to the existence of some duality relation.  In
particular, in the correspondent CG formulation of the Potts model,
this duality is expressed in terms of an electric-magnetic duality
transformation, also called T-duality in the literature
\cite{Gruzberg}. The T-duality relate the descriptions of the dilute
and dense phase of the correspondent $O(n)$ model and it is at the
basis of the Duplantier duality \cite{Duplantier2}.

The natural question for the $Z_N$ model is then to see which are the
relevant clusters. The answer, that we will explain in great details
in this paper, is that i) the spin clusters percolate at the
critical point and the associated exponents do not correspond to the
corresponding model. ii) the FK clusters do {\bf not} percolate at the
critical point.

We will provide some details of the cluster algorithms that we
employed in this study. Cluster algorithms have been first employed on
the Potts model. The Potts model for any number of states is a two
level local energy model on a lattice. The energy between two spins is
either zero or a fixed value ($\beta$). Then the clusters are defined,
for a fixed configuration of spins, as a problem of percolation. On
all spin clusters which are build as neighboring spins taking
the same value and connected with a term of energy $\beta$, one
connects each pairs of spins with a probability $p=1-e^{-\beta}$. The
resulting clusters of connected spins are the Fortuin Kastelyn
clusters which are used to build the dynamics of the model but also to
measure observables like the magnetisation or the magnetic
susceptibility.

For the $Z(N)$ parafermionic theory that we will consider here, the
situation is more complicated. The local energy can take more than two
values for $N \geq 4$ and a direct consequence is that the generalised
clusters can connect spins with different values. Moreover, while for
the Potts models it was possible to defined some quantities as the
size of some FK clusters, this will not be the case here.

\section{Definitions}
One consider a model of spins variables $S_i$ which can take $N$
values, $S_i = 1, \cdots, N$ and are located on a square lattice of
linear size $L$ with periodic boundary conditions on both directions.

We consider the model defined on a square lattice with spin variable
$S_j=\exp{i 2\pi/N n(j)}$ at each site $j$ taking $N$ possible
values, $n(j)=0,1,\cdots,N-1$. The most general $Z_N$ invariant spin
model with nearest-neighbor interactions is defined by the reduced
Hamiltonian \cite{Zama_lat,Dotsi_lat}:
\begin{equation}
H[n]=-\sum_{m=1}^{\lfloor N/2 \rfloor} J_{m}\left[\cos \left(\frac{2\pi m n}{N}\right)-1\right],
\label{reducedH}
\end{equation}
where $\lfloor N/2 \rfloor$ denotes the integer part of $N/2$. The
associated partition function reads:
\begin{equation}
Z=\sum_{\{S\}}\exp\left[-\beta \sum_{<ij>} H[n(i)-n(j)]\right] \; .
\label{partition1}
\end{equation}
For $J_m=J$, for all $m$, one recovers the $N-$state Potts model,
invariant under a permutational $S_N$ symmetry while the case
$J_m=J\delta_{m,1}$ defines the clock model \cite{Potts_Clock}. For
$N=2$ and $N=3$ these models coincide with the Ising and the
three-state Potts model respectively, while the case $N=4$ is
isomorphic to the Ashkin-Teller model \cite{Ashkin,Lin}. Defining the
Boltzmann weights:
\begin{equation}
x_n=\exp\left[-\beta H(n)\right], \quad n=0,1,\cdots,N-1 \; ,
\end{equation}
the most general $Z_N$ spin model is then described by $\lfloor
N/2\rfloor$ independent parameters $x_n$ as $x_0=1$ and
$x_n=x_{N-n}$. The general properties of these models for $N=5,6,7$
have been studied long time ago (see e.g. \cite{Alcaraz2} and
references therein). The associated phase diagrams turn out to be
particularly rich as they contain in general first-order, second-order
and infinite-order phase transitions.  For all the $Z_N$ spin models
it is possible to construct a duality transformation (Kramers-Wannier
duality). In the self-dual subspace of
(\ref{reducedH})-(\ref{partition1}), which also contains the Potts and
the clock model, the $Z_N$ spin model are critical and completely
integrable at the points \cite{Zama_lat2, Alcaraz}~:
\begin{eqnarray}
x^{*}_0= 1 \; ; \; x^{*}_n&=&  \prod_{k=0}^{n-1} \frac{\sin \left(\frac{\pi k}{N}+\frac{\pi}{4
    N}\right)}{\sin \left(\frac{\pi (k+1)}{N}-\frac{\pi}{4 N}\right)} \; .
\label{integrablecond}
\end{eqnarray}
There is strong evidence that the self-dual critical points
(\ref{integrablecond}), referred usually as Fateev-Zamolodchikov 
points, are described in the continuum limit by $Z(N)$
parafermionic theories \cite{Alcaraz3}.  Very recently, a further
strong support for this picture has been given in \cite{Rajabpour}
where the lattice candidates for the chiral currents generating the
$Z_N$ symmetry of the continuum model has been constructed.

\section{Cluster algorithm}

In this section, we explain how we can generalise the notion of FK
clusters to the case of the $Z_N$ spin model. We
will consider configurations on a square lattice of linear size $L$
with periodic boundary conditions for which we need to generate
independant samples. The most convenient way to generate these samples
is to use a cluster algorithm. The most effective cluster algorithm
for discrete spin models is the Wolff~\cite{Wolff} algorithm which is
based on the construction of the Fortyuin Kastelyn~\cite{FK} clusters. We first recall 
how this algorithm works in the simple case of the $N$-states Potts model. Starting from
\beq
Z=\sum_{\{S\}} e^{\beta \sum_{<i,j>} \delta_{S_i S_j}} \; ,
\eeq
where the first sum $\{S\}$ is on all the spins $S_i=1,\cdots,N$ while the
second sum $<i,j>$ is on the first neighbor spins on the lattice, one easily gets
\beq
Z=\sum_{\{S\}} \prod_{<i,j>} e^{\beta \delta_{S_i S_j}} 
=(e^{\beta})^M \sum_{\{S\}} \prod_{<i,j>} ((1-e^{-\beta}) 
\delta_{S_i S_j} + e^{-\beta})  \; ,
\eeq
with $M$ the total number of bonds on the lattice. Defining
$p=1-e^{-\beta}$, the partition function is
\beq
Z=(e^{\beta})^M \sum_{\{S\}} \prod_{<i,j>} ( p \delta_{S_i S_j} + (1-p))  \; .
\label{decompositionQ}
\eeq
$ $From there, one can read the rules to build the FK clusters. In a
given configuration, for two neighbouring spins $i$ and $j$ such that
$S_i=S_j$, one will put a bond with probability $p$ and no bond with
probability $1-p$.

In the case of the $Z_N$ model the situation is a little bit more
complicated. The partition function (\ref{partition1}) can be written as 
\beq 
Z_N =\sum_{\{S\}} \prod_{<i,j>} 
(x^{*}_0)^{\delta_{n(i), n(j)}} 
(x^{*}_1)^{\delta_{n(i), n(j)\pm 1}} 
\cdots 
(x^{*}_{[N/2]})^{\delta_{n(i), n(j) \pm [N/2]}} \; ,
\eeq
the delta function being defined modulo $N$, ie $\delta_{a,b} = 1
$ if $a \equiv b \; (N)$. A decomposition similar to the one of
(\ref{decompositionQ}) is
\bea
Z_N &=& \sum_{\{S\}} \prod_{<i,j>} ( 1-x^{*}_{[N/2]}) \delta_{n(i),n(j)} 
+ (x^{*}_1-x^{*}_{[N/2]}) \delta_{n(i), n(j)\pm 1} \nn \\
&& + \cdots + (x^{*}_{[N/2]-1}-x^{*}_{[N/2]}) \delta_{n(i),n(j)\pm ([N/2]-1)} 
+ x^{*}_{[N/2]}  \; .
\eea
Note that due to the definition of the $x^{*}_i$, cf
eq.(\ref{integrablecond}), the $x^{*}_i$'s will be ordered and
positives, $1=x^{*}_0 > x^{*}_1 > \cdots > x^{*}_{[N/2]} > 0$. From
there, one can read the construction of the generalised FK
clusters. For each pair of neighbouring spins $S_i$ and
$S_j$, one will put a bond with probability
\beq
p_{|n(i)-n(j)|}={x^{*}_{|n(i)-n(j)|} - x^{*}_{[N/2]}\over
  x^{*}_{|n(i)-n(j)|}}
\eeq
and no bond with probability $1-p_{|n(i)-n(j)|}$. These FK clusters
will be used to construct a cluster algorithm of Wolff
type~\cite{Wolff}. A lattice update consists in selecting one spin in
the lattice at a random location then building the FK cluster
containing this spin and then changing the color of this cluster by
changing each spin of the lattice as $S_i \rightarrow S_i + j (N)$
with a random value $1 \leq j \leq N-1$.

It is important to note that for the $Z_N$ models, the FK cluster will
connect spins with different values which is not the case for the
Potts model. One important consequence is that the resulting FK
clusters can not be associated directly to some physical quantities
like it was the case for the Potts models. For these models, the FK
clusters are the basic ingredient for building an update algorithm but
they also encode all the informations associated to the critical
behavior of the model under consideration. For example, the average
size of a cluster build from a random site (Wolff algorithm) is equal
to the magnetic susceptibility. Or the average size of the largest FK
cluster divided by the volume is equal to the magnetization of the
system. Or the two point spin-spin correlation function is equal to
the probability that the two spins are in the same FK cluster. All
these relation can not be valid any more in the case that we consider
here. Still similar quantities can be defined. For example, if one
defines for each cluster $k$
\beq
\rho_k = |\sum_{i} < e^{2 i \pi n(i) \over N} > | \; ,
\label{mag1}
\eeq
the sum being restricted to all the spins in the cluster $k$, then the
magnetisation is associated to the maximum $\rho_k$ along all the
clusters. This is a simpler generalisation of the Potts model for
which each FK cluster contains only spins with identical sign, thus in
that case $\rho_k$ is the volume of the FK cluster. We numerically
compared the quantity $mag_1(L) = (\max (\rho_k)/ L^2)$ with the real
magnetisation obtained as a weighted sum on all the lattice
\beq
mag(L) = {1\over L^2} |\sum_{i=1,L^2} < e^{2 i \pi n(i) \over N} > | \; , 
\label{mag2}
\eeq
the agreement being perfect for both the $Z_4$  and the $Z_5$ spin models. 
\begin{figure}
\begin{center}
\epsfxsize=400pt{\epsffile{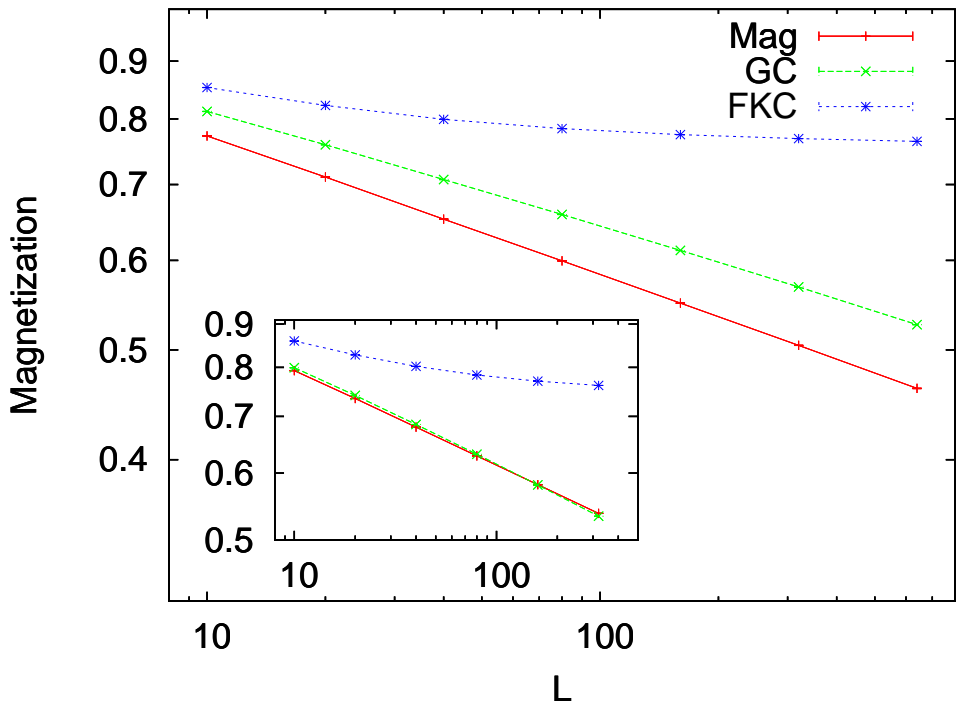}}
\end{center}
\caption{
Main panel: Magnetization vs. $L$ for the $Z_4$ spin model. The magnetization
has been computed in three different ways: from eq.~(\ref{mag1}) and
from the average of the largest spin and FK clusters. In the inset,
the same plot for $Z_5$ spin model.
\label{PlotM}
}
\end{figure}
In the main panel of Fig.\ref{PlotM}, we plot for the $Z_4$ spin
model, the magnetisation obtained from eq.~(\ref{mag1}) which is
compared to the value of the average largest FK cluster and the
average largest spin cluster (in both cases divided by
$L^2$). The scaling is perfect for the magnetization given by
eq.~(\ref{mag1}) with an exponent in very good agreement with the
expected one, $\beta/\nu=1/8$ \cite{Zama_lat2}. For the spin cluster, we also
observe a good scaling but with stronger finite size corrections. Due
to these corrections, it is difficult to give a definite exponent
associated to the spin clusters, we obtain for the largest
sizes $\beta/\nu = 0.110(1)$. Since this value is increasing with the
size, one can speculate that in the infinite size limit this value
will converge towards the magnetic exponent $\beta/\nu$. We also
observe that the largest FK cluster will occupy a finite fraction of
the lattice in the large size limit, which corresponds to the case
where the percolation threshold has been exceeded. We will come back
on this point in the next section.

In the inset of Fig.\ref{PlotM}, we show similar data for the $Z_5$
spin model. We also obtain an excellent agreement between the exponent
obtained from eq.~(\ref{mag1}), the real magnetisation
eq.~(\ref{mag2}) and the exact result $\beta/\nu = 4/35$ \cite{Zama_lat2}. We see that
the spin cluster exponent is again affected by strong finite
size effects and it will become larger than the magnetic exponent
already for the simulated sizes (note the crossing between this curve
and the one corresponding to the magnetization from
eq.~(\ref{mag1})). As for the $Z_4$ case, the largest FK cluster do
not present a scaling law at the critical point.
\begin{figure}
\begin{center}
\epsfxsize=400pt{\epsffile{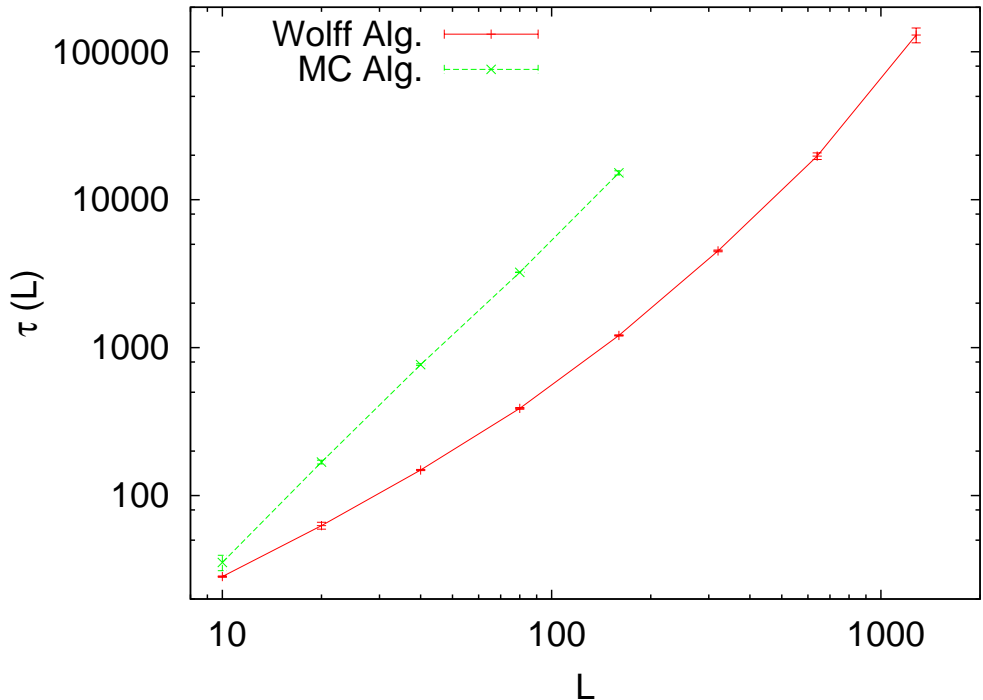}}
\end{center}
\caption{
Autocorrelation time vs. $L$ for the $Z_4$ spin model. We show the data for the Wolff 
algorithm as well as for ordinary Monte Carlo updates. The Wolff algorithm show a much better efficiency
in the range of lattice sizes explored.
\label{Pauto}
}
\end{figure}

Even if we expect that the FK clusters are not the natural object to
compute the critical exponents of the $Z_N$ spin models, we can still
use them to build cluster algorithms. In Fig.~\ref{Pauto} we plot the
autocorrelation time for the Wolff algorithm in the $Z_4$ spin model
and we compare it to the autocorrelation time for standard heat bath
Monte Carlo.  We observe that the Wolff algorithm is much more
effective than the Monte Carlo one. The autocorrelation time for the
Monte Carlo algorithm scales as $\tau(L) \simeq L^{z_{MC}}$ with
$z_{MC} = 2.1(1)$. The dynamical exponent $z$ is much smaller for the
Wolff algorithm at small sizes ($z_W \simeq 1.2(1)$) but then it
increase for larger sizes.  This effect will be explained in the next
section. For the sizes that we can simulate, $L \leq 1280$, the Wolff
algorithm will always be more efficient than Monte Carlo. This is also
the case for the $Z_5$ spin model.

\section{Percolation and critical properties}

In this section, we present results for the properties of both
spin and FK clusters in the $Z_4$ and $Z_5$ spin models. We
show that the distribution of cluster lengths in the critical point is a
power law for spin clusters but not for FK clusters.
Furthermore we show that the spin clusters percolate at the
critical point, while the FK clusters do not. We also perform a first
determination of the fractal dimension of the spin clusters by
using the exponent associate to the distribution of cluster lengths.
\begin{figure}
\begin{center}
\epsfxsize=400pt{\epsffile{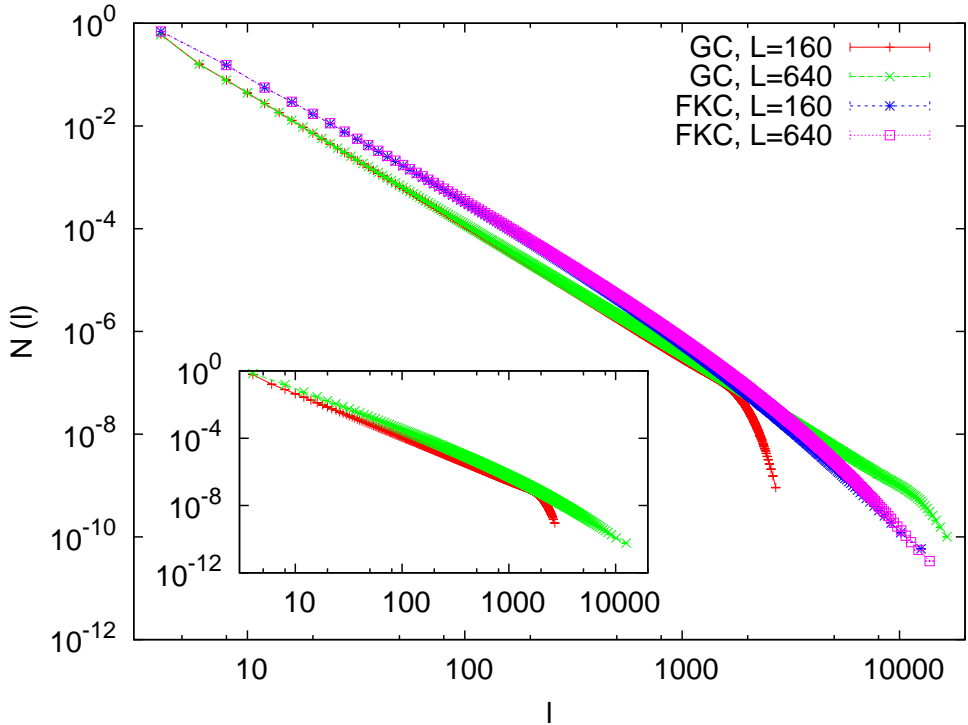}}
\end{center}
\caption{
Distribution of cluster lengths for the
$Z_4$ spin model and different lattices sizes. While the spin clusters 
show a nice power law distribution, the FK cluster distribution falls away from a power law.
In the inset, we show similar data for the $Z_5$ spin model and $L=160$.
\label{PlotD}
}
\end{figure}

First, we consider the distribution of the length \footnote{As
  explained in the next section, there exist two natural ways to
  define a length.  Both methods converge to the same result in the
  large size limit.} of contours for the finite size clusters at
the critical point. In Fig.\ref{PlotD}, we present this distribution
for both spin and FK type of clusters in the $Z_4$ spin model for
$L=160$ and $640$. For both lattice sizes, we observe a nice scaling
for the spin clusters with a distribution which is well
described by a power law $N(l) \simeq l^{-\tau_g}$ characteristic of a
percolation critical point~\cite{ Stauffer} with $\tau_g \simeq
2.5(1)$.  We expect that $\tau_g$ is related to the fractal dimension
by the following relation $d_f = 2/(\tau_g-1) \simeq 1.33(10)$ which
is of the same order of what is measured in the SLE context
\cite{MarcoRaoul,MarcoRaoul2}. In the next section, we will present
more precise measurements for the fractal dimensions of the spin
clusters. For the FK clusters, it is clear that the scaling is not
satisfied. The distribution is better described by $N(l) \simeq
l^{-\tau_{fk}} \exp{(-l / \xi)}$ with some finite correlation length
$\xi$ with a value of order 2500 lattice units for the $Z_4$ spin model.
\begin{figure}
\begin{center}
\epsfxsize=400pt{\epsffile{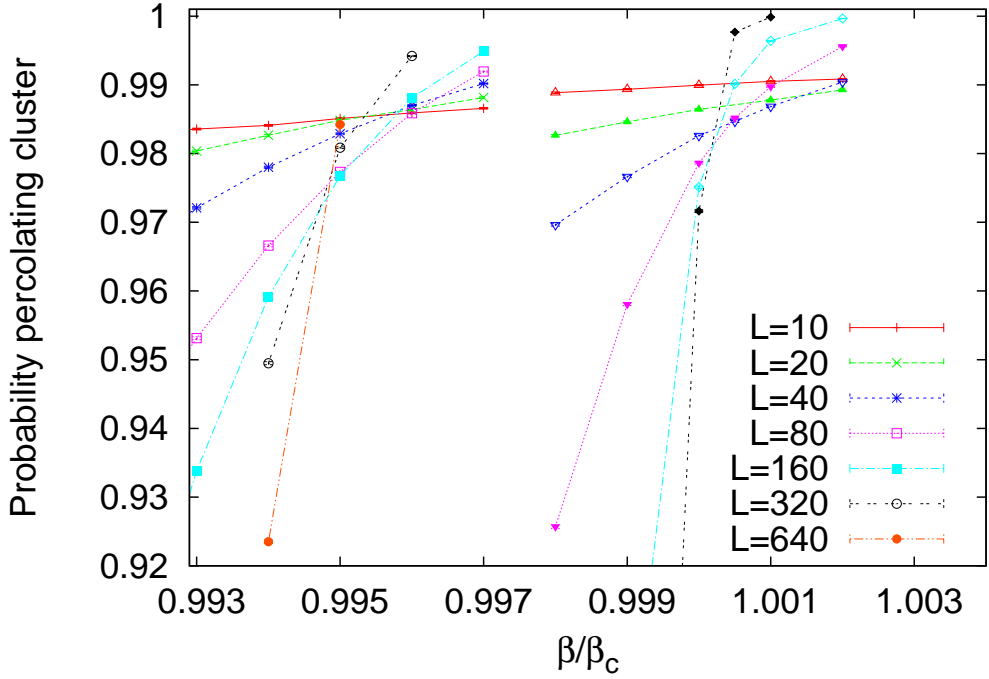}}
\end{center}
\caption{
Percolation test for $Z_4$ spin model. The right part corresponds to the spin clusters, which
percolate right at the critical point. The left corresponds to the FK clusters, which
do not percolate at this point.
\label{PlotP}
}
\end{figure}
\begin{figure}
\begin{center}
\epsfxsize=400pt{\epsffile{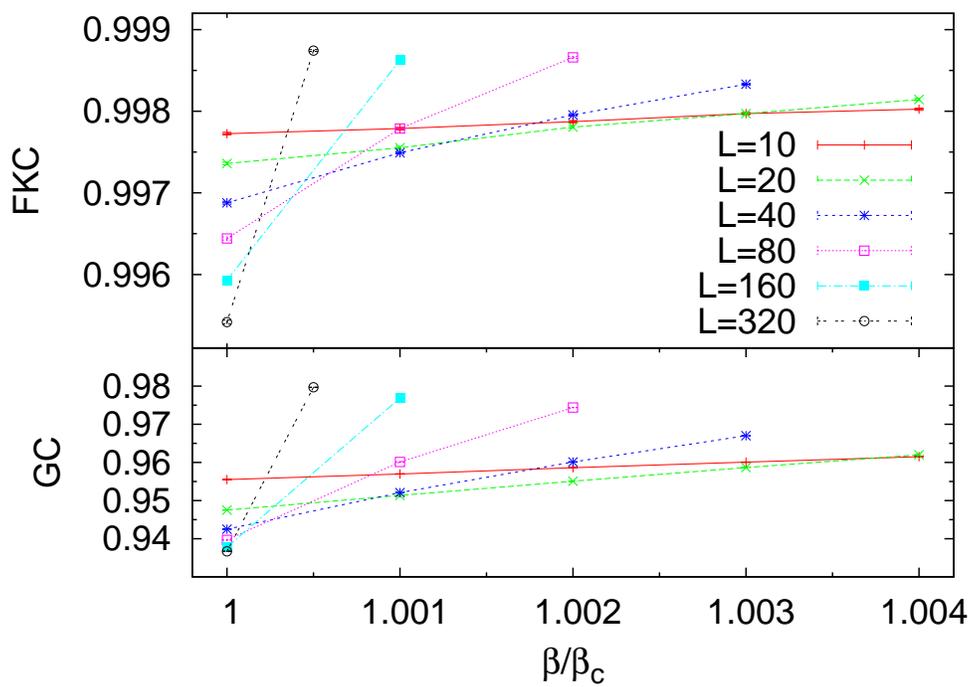}}
\end{center}
\caption{
Percolation test for $3$-states Potts model. The upper part corresponds to the FK clusters, while 
the lower part corresponds to the spin clusters. Both types of clusters 
percolate at the critical point.
\label{Plotc}                                                  
}
\end{figure}

\begin{figure}
\begin{center}
\epsfxsize=400pt{\epsffile{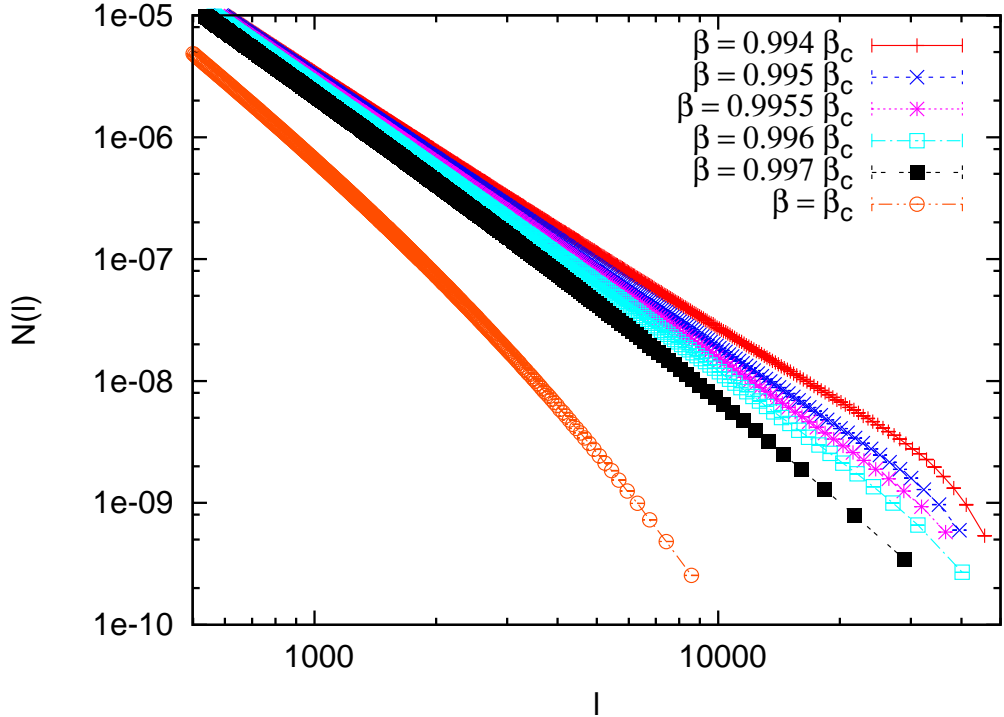}}
\end{center}
\caption{
Distribution of FK cluster lengths for the $Z_4$ spin model and $L=320$ for different values
of the parameter $\beta$ very near the critical point.
\label{Plotb}
}
\end{figure}

This provides a first evidence that the FK clusters do not percolate
at the critical point of the $Z_4$ spin model. This is confirmed in
Fig.\ref{PlotP}, where we present the probability of having a
percolating cluster vs. the ratio $\beta/\beta_c$ (which measures the
distance to the critical point), for both the spin and the FK
clusters. The probability is computed for increasing lattice
sizes. Converging crossing points indicate a critical point. This is
clearly observed for the spin clusters with a critical point
close to $\beta=\beta_c$. For the FK clusters we do not observe a
clear convergency and the lines cross around $\beta \simeq 0.995
\beta_c$. This corresponds to the finite correlation length previously
observed for the distribution of FK clusters length. To convince the
reader that such a small deviation, $|\beta-\beta_c|/\beta_c \simeq
0.005$, is in fact important, we show in Fig.\ref{Plotc} a comparative
plot for the $3$-states Potts model.  For this model, both types of
clusters percolate at $\beta_c$. For the larger lattice size, we see
that the deviation is near two order of magnitude smaller than it was
for the $Z_4$ spin model. The percolation value $\beta \simeq 0.995
\beta_c$ is also confirmed in Fig.~\ref{Plotb} where we plot the
distribution $N(l)$ of the FK clusters for various value of $\beta$
and for $L=320$. We observe a nice power law close to $\beta = 0.995
\beta_c$.

In the case of the $Z_5$ spin model we obtain similar
results.  The distribution of spin cluster lengths shows a
power law scaling, while the length distribution of FK clusters
presents a finite correlation length $\xi$ with a value of order 5000
lattice units. See the inset of Fig.~\ref{PlotD}. Furthermore,
spin clusters percolate right at the critical point
$\beta=\beta_c$, while FK clusters percolate at $\beta = 0.9975
\beta_c$.

\section{Fractal dimensions in the bulk}

In this section we present a more accurately computation for the fractal
dimensions of the spin clusters for both $Z_4$ and $Z_5$ spin models. 
\begin{figure}
\begin{center}
\epsfxsize=400pt{\epsffile{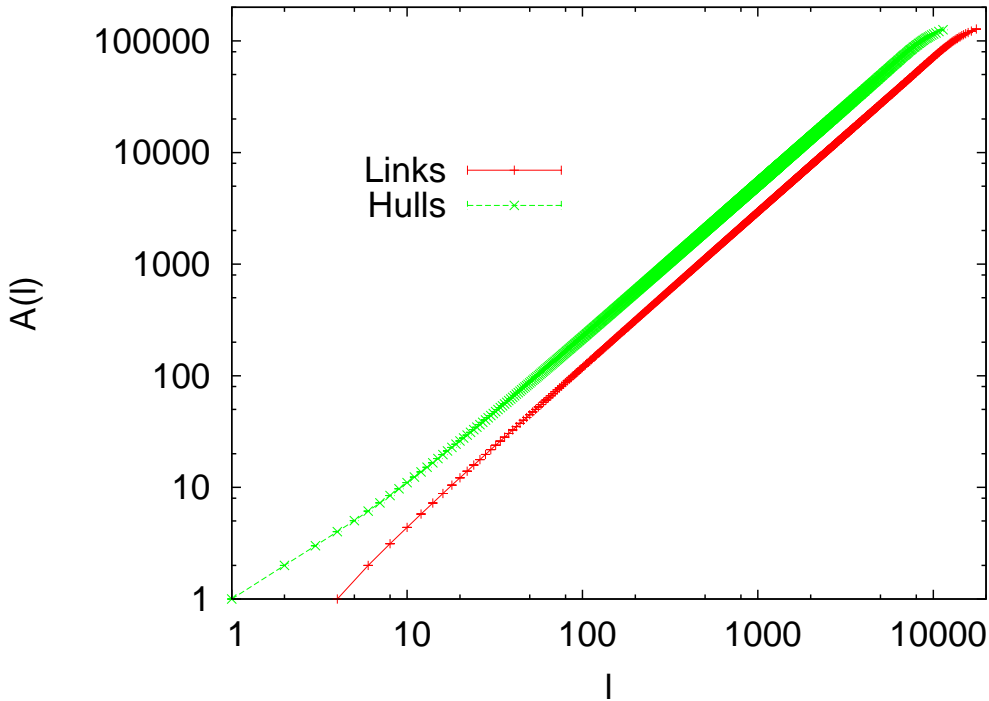}}
\end{center}
\caption{
$A(l)$ vs. $l$ for the two definitions of the length (links and hulls, see text for details) on the
spin clusters of the $Z_5$ spin model.
\label{Plot1}
}
\end{figure}

As explained in the previous section, the fractal dimension can be
obtained from the distribution $N(l) \simeq l^{-\tau}$ via the
relation $d_f=2/{(\tau-1)}$. This method turns out not to be very precise 
since there exist very strong finite size corrections in the determination of $\tau$. 
Here we present another method which
provides a better precision by computing the average area of the
cluster as a function of the interface length around the cluster. We
will consider two different definitions of the cluster interface
length. The first one, which we will call the {\it link length}
corresponds to the number of bonds which are broken around the
cluster. For example, for one isolated spin, the length is
$l_l=4$. The second definition, which we will call the {\it hull
  length}, counts the number of spins on the border of the
cluster. For an isolated spin, one has $l_h=1$.

The fractal dimension is defined as $l = R^{d_f}$ with $R$ the radius
of gyration of the cluster. A more direct measurement is given by
computing the average area as a power law of $l$ with $A(l) = R^2
\simeq l^{2/d_f}$. In Fig.~\ref{Plot1} we show the data for the $Z_5$ 
spin model. In this
plot, we present $A(l)$ for two definitions of $l$, the hull and the
link one. We get a nice scaling law over
a large range of $l$'s. The asymptotic limit is similar for the two
definitions of lengths (hulls and links). A fit of the data gives a value of $d_f \simeq
1.44 (1)$, but such a fit does not provide a good precision since it
is very difficult to take in account the small and large size
corrections. Note that there is a bending in both of these curves for small sizes. 
These bendings, which are due to small size corrections, are in opposite directions for the 
two definitions of lengths that we employ. This fact will be very useful for the extraction 
of a precise fractal dimension and motivate the measurement of the two lengths.
\begin{figure}
\begin{center}
\epsfxsize=400pt{\epsffile{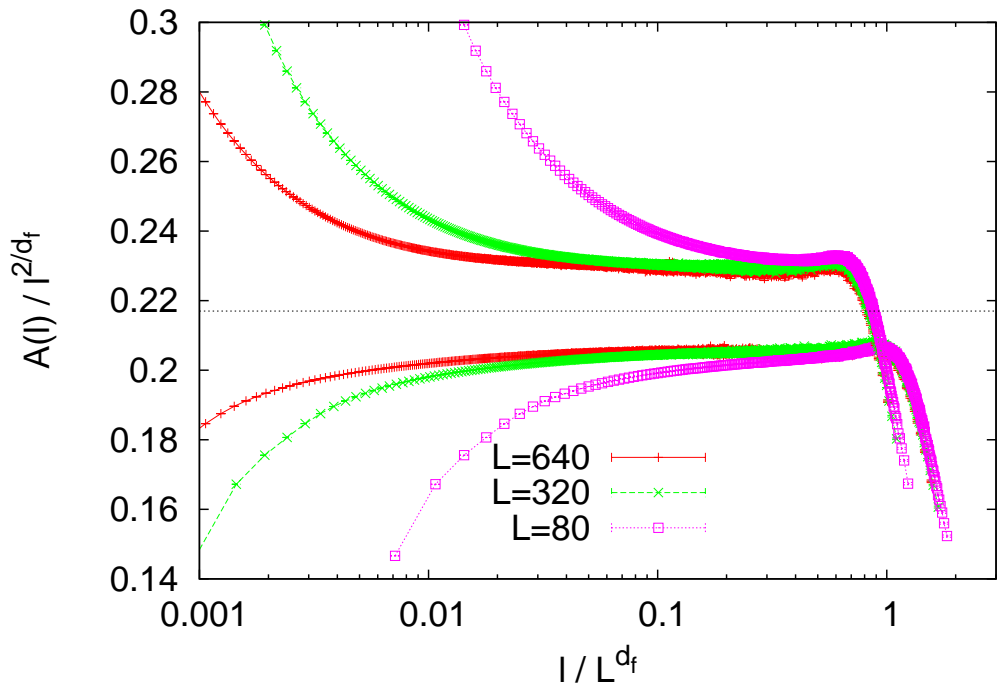}}
\end{center}
\caption{
Rescaling for the $Z_5$ spin model and different system sizes with $d_f=1.446$.
The upper curves correspond to the hull lengths with the lower ones correspond
to the link lengths.
\label{Plot2}
}
\end{figure}

A better estimate is obtained in the following way: in
Fig.~\ref{Plot2} we show a similar plot after a rescaling $l
\rightarrow l/L^{d_f}$ and $A(l) \rightarrow A(l)/l^{2/d_f}$.  The
rescaling is motivated by the fact that $L^{d_f}$ corresponds the
length of a cluster who fills the lattice, {\it i.e.} $A
=(L^{d_f})^{2/d_f} = L^2$. We observe a collapse for the large size
clusters. There still exists strong finite size corrections, for both
small and large cluster lengths, but we see that a plateau appears
which correspond to the region where scaling works.
%Again we observe
%that the small sizes corrections are in opposite directions for the
%two lengths considered which helps in the determination of $d_f$. 
The optimal values are $d_f=1.450(2)$ for hulls and $d_f=1.444(2)$ for
links.
\begin{figure}
\begin{center}
\epsfysize=130pt{\epsffile{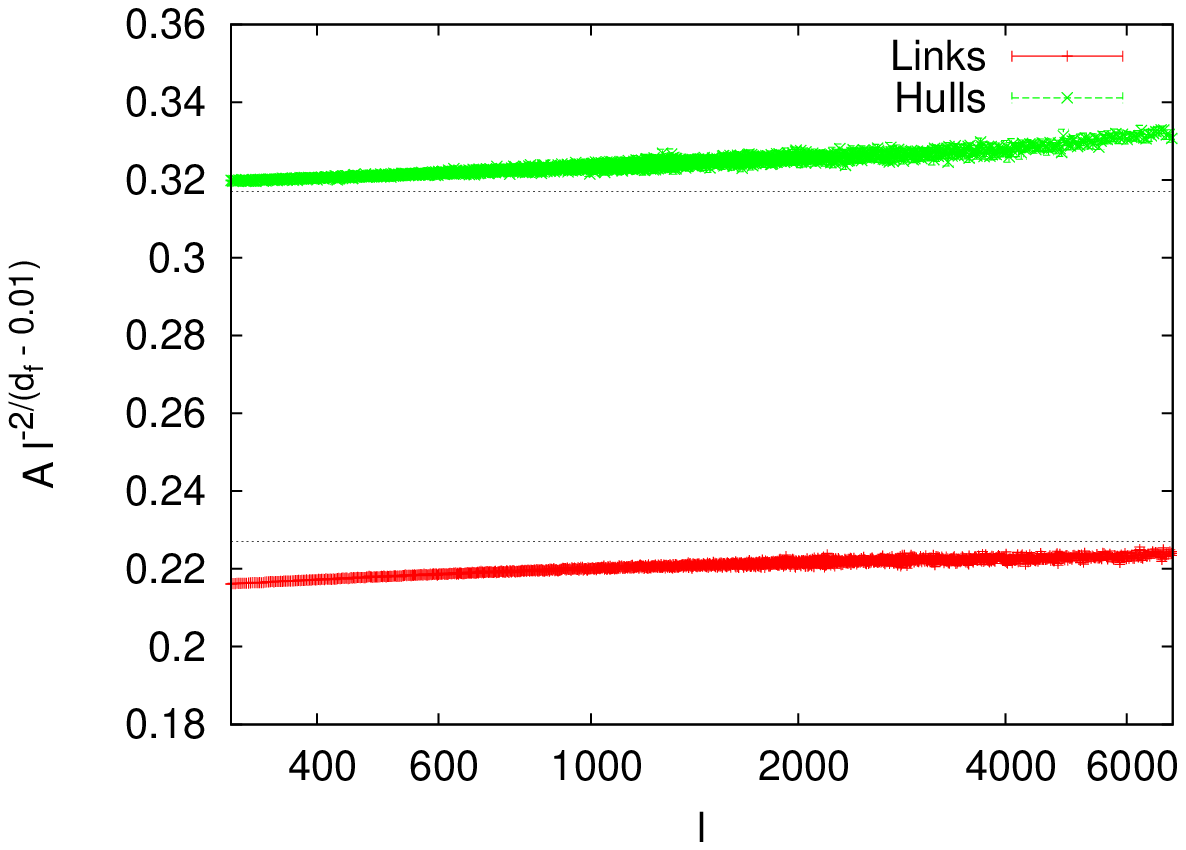}}
\epsfysize=130pt{\epsffile{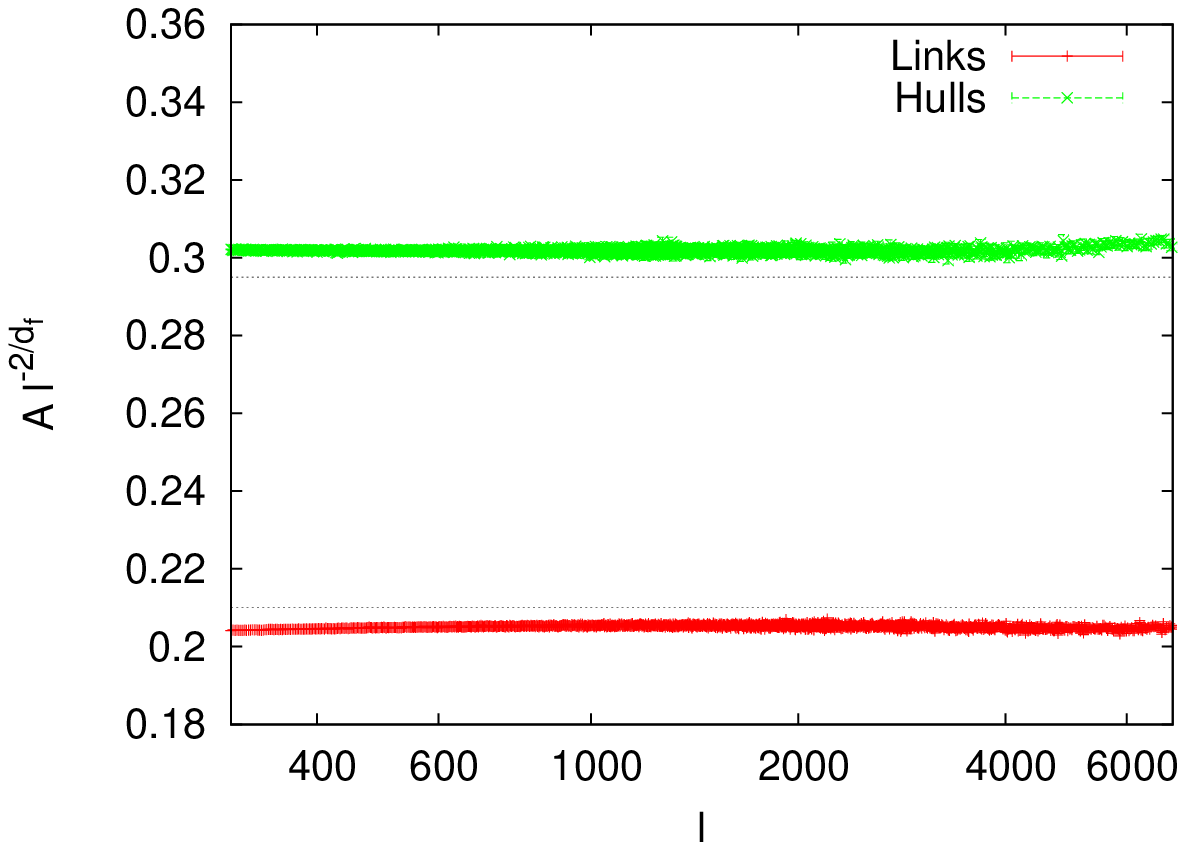}}
\epsfysize=130pt{\epsffile{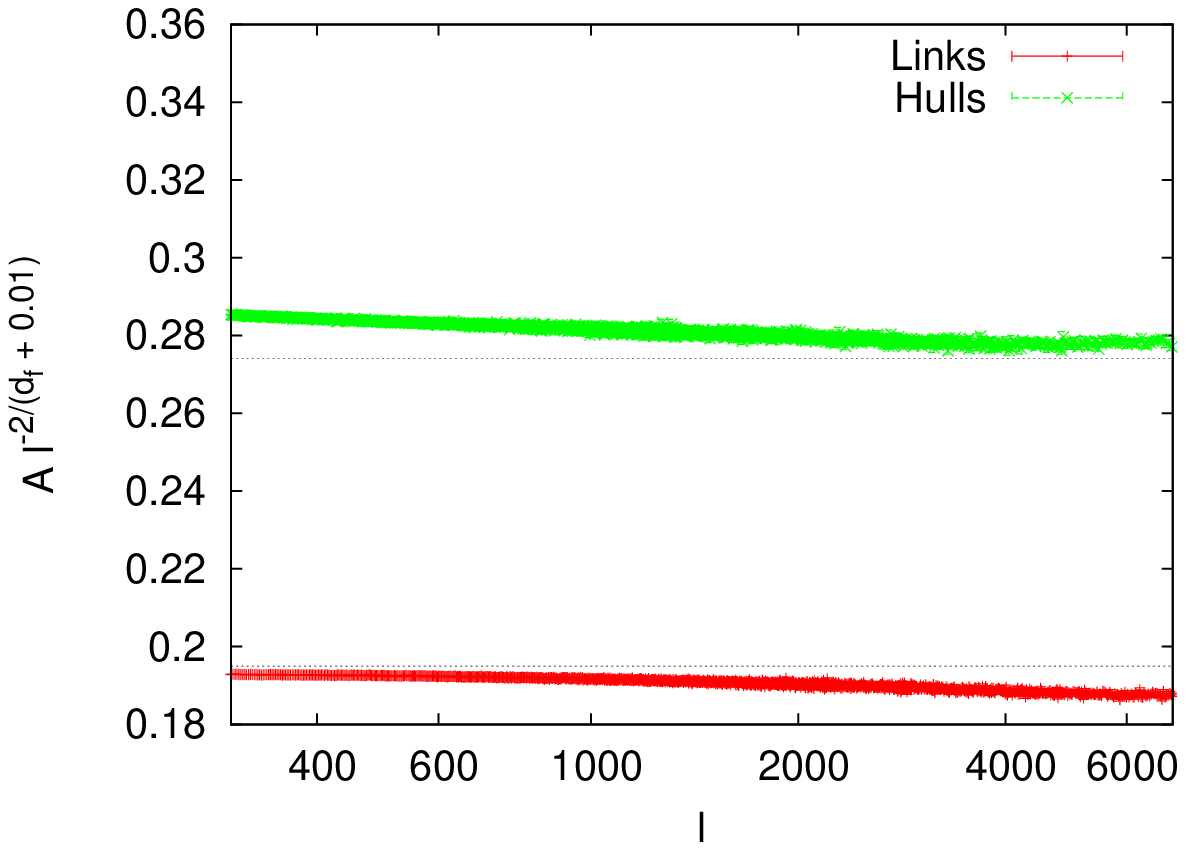}}
\end{center}
\caption{
Comparative with three values $d_f-0.01$, $d_f$ and $d_f+0.01$ 
%for both definitions of cluster length and
for $L=640$ in the $Z_5$ spin model with $d_f=1.450$ for the hull cluster length and $d_f=1.444$ 
for the link cluster length. This plot show
the accuracy on the determination of the fractal dimension.
\label{Plot3}
}
\end{figure}
In Fig.~\ref{Plot3} we check the accuracy of our estimation for the
$Z_5$ fractal dimension.  By plotting ${A /l^{2/d_f}} $ vs. $l$, the
good value of the fractal dimension should correspond to straight and
perfectly horizontal lines. As shown in the figure, this is obtained
for an optimal value of $d_f=1.446(2)$ (which correspond to the
average of hull and link fractal dimension).

\begin{figure}
\begin{center}
\epsfxsize=400pt{\epsffile{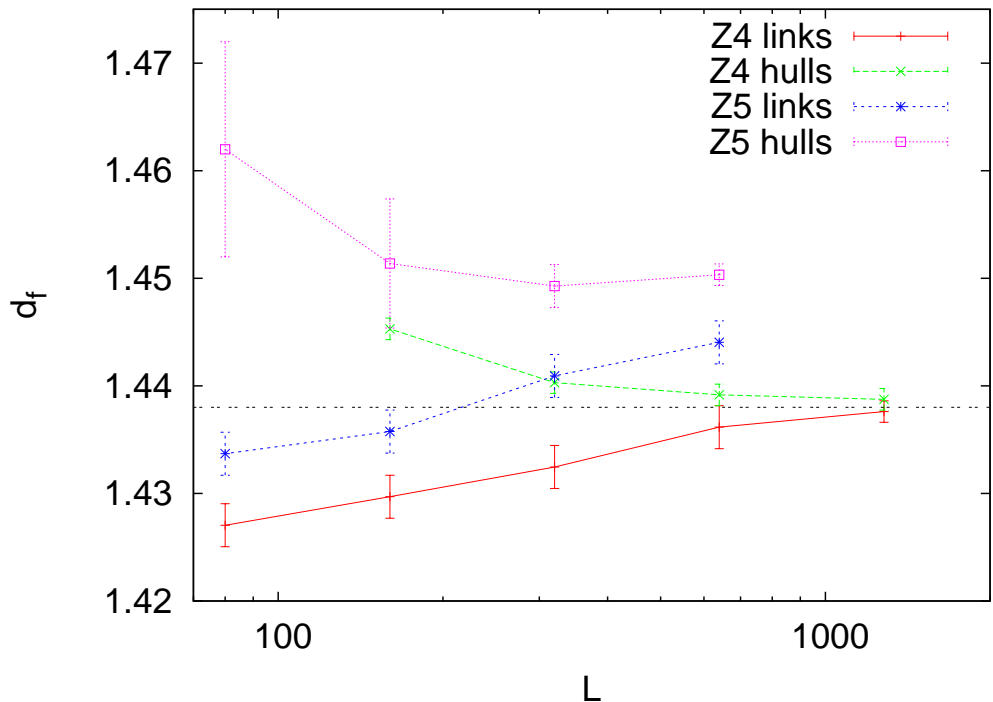}}
\end{center}
\caption{
Estimations of the fractal exponents for both hull and link lengths, in the $Z_4$ and $Z_5$ spin
models as a function of the lattice size.
\label{Plot4}
}
\end{figure}

For the $Z_4$ spin model, we can perform similar
measurements. We show in Fig.~\ref{Plot4}, values obtained for $d_f$ for
$Z_4$ and $Z_5$ and for both definitions of the length. This figure contains the
main results of our work. As a final result for the fractal dimension of the spin clusters,
we obtain $d_f=1.438(2)$ for the $Z_4$ spin model and $d_f=1.446(2)$ for the $Z_5$ spin model.
It is important to note that these
values are different from the ones proposed by one of the authors in
\cite{Raoul}. Here a particular interface associated to some (conformal) boundary condition
was considered. The corresponding theoretical predictions were based on the conformal properties (in particular a two level null vector conditions) of the boundary condition changing operator and thus were explicetly dependent on the boundary conditions which generate such interface. One has to take into account that, in the case of non-minimal $Z_N$ theories, the classification of conformal boundary conditions becomes more rich. For instance, differentely from the $Z_3$ case, there are for the  $Z_4$ and $Z_5$ models two different mixed boundary conditions related to operators satisfying a two-level null vector condition.      
A detailed study, which will be present in a future publication \cite{MarcoRaoul2}, shows that the fractal dimension of some interfaces associated to different boundary conditions are different. Still the fractal dimension obtained
for interfaces to some type of boundary conditions are in good agreement with the one determined 
here in the bulk \cite{MarcoRaoul2}.
We point out also that the fractal dimensions of the bulk spin cluster boundaries  obtained here do not correspond to any of the values (or their dual) proposed in \cite{Rajapour} and concerning interfaces in $O(n)$ loop models related to parafermionic theories.   

\section{Summary and conclusions}
In this paper we studied by Monte Carlo methods the spin
properties of the $Z_N$ spin model.  The samples were generated by
using a cluster algorithm which generalize the notion of FK clusters
to the case of the $Z_N$ spin model.  For $N\geq 4$ the FK clusters
will in general connect spins with different values. This is not the
case for $N=2$ and $N=3$, respectively the Ising and three-states
Potts model.  The cluster algorithm allows to track both spin
and FK clusters and the distribution of all the finite closed
spin and FK clusters can be studied.
  
In particular we have shown that the spin clusters percolate at
the critical point and the associated exponents do not correspond to
the exponents given by the unitary Kac table of the associated $Z_N$
parafermionic field theory. Note that this is also true for the
spin clusters for $N=2,3$.  We have determined the fractal
dimension of the boundaries (interface) of the spin clusters. By
measuring the distribution in size and area of the spin
clusters, we determined the fractal dimension $d_f = 1.438(2)$ for
$N=4$ and $d_f=1.446(2)$ for $N=5$. These
values are different from the ones proposed by one of the authors in
\cite{Raoul} for SLE interfaces in parafermionic theories 
and measured numerically for some particular type of boundary
condition in \cite{MarcoRaoul}. Still the fractal dimension obtained
by numerically studying interfaces related to certain different types
of boundary conditions are in good agreement with the one determined 
here in the bulk \cite{MarcoRaoul2}.  

We have also shown that, although the cluster (Wolff) algorithm
show a much better efficiency in the range of system lengths studied,
the FK clusters do {\bf not} percolate at the critical point for
$N\geq 4$. This is the reason why we computed the fractal dimension 
only for the spin clusters.

The results we obtained point out important differences in the
behavior of the spin and FK clusters between the case $N=2,3$,
where the system can be described by a minimal CFT model, and the case
$N\geq 4$, described in the continuum limit by an extended CFT. This
can be traced back to the fact that, for $N\geq 4$ the internal
$Z_N$ degrees of freedom play a fundamental role. This calls for
further analytical studies of the bulk geometric properties of the
parafermionic theories.  One way to tackle this problem would be to
provide a Coulomb gas description for parafermionic theories which
would allow to identify the operators associated to the geometric
interface and to compute the associated fractal dimensions.

\noindent{\large\bf Acknowledgments} 

\vspace*{0.7 true cm} We thanks Benoit Estienne for useful
discussions. This work has been done in part when one of the authors
(RS) was guest of the Galielo Galilei Institute in Florence, whose
hospitality is kindly acknowledged.

\noindent

\newpage
%\small


\begin{thebibliography}{99}
\bibitem{Duplantier} see for instance B. Duplantier, 
\newblock Les  Houches 2005 Lecture Notes, Session LXXXIII, 101,
Elsevier  (2006), math-ph/0608053

\bibitem{Nienhuis_CG} B.~Nienhuis,
\newblock in  {\it Phase Transitions
and Critical Phenomena}, edited 
by C. Domb and J. L. Lebowitz (Academic, London, 1987), Vol. 11, p.1.

\bibitem{Saleur} H.~Saleur and B.~Duplantier,
\newblock {\it Phys. Rev. Lett.}~{\bf 58}, 2325 (1987).

\bibitem{Duplantier2} B.~Duplantier, 
\newblock {\it Phys. Rev. Lett.}~{\bf 84}, 1363 (2000).

\bibitem{Walter} W.~Kager and B.~ Nienhuis, 
\newblock {\it J. Stat. Phys.}~{\bf 115}, 1149 (2004).

\bibitem{Cardy_review} J.~Cardy,
\newblock {\it Annals Phys.}~{\bf 318}, 81 (2005).

\bibitem{Bernard_review} M.~Bauer and D.~Bernard, 
\newblock {\it Phys. Rept.}~{\bf 432}, 115 (2006).

\bibitem{Bernard_connection1} M.~Bauer and D.~Bernard,
\newblock {\it Comm. Math. Phys.}~{\bf 239}, 493 (2003).

\bibitem{Bernard_connection2} M.~Bauer and D.~Bernard, \newblock
  {\it Phys. Lett.}~{\bf B543}, 135 (2002).

\bibitem{Bernard_connection3} M.~Bauer and D.~Bernard, \newblock
 {\it Phys. Lett.}~{\bf B557}, 309 (2003).

\bibitem{Affleck} I.~Affleck and F.~D.~M.~Haldane,
 \newblock {\it Phys. Rev.}~{\bf B36}, 5291 (1987).

\bibitem{Ludwig} A.~W.~W.\ Ludwig,
\newblock {\it Nucl.\ Phys.}~{\bf B285}, 97 (1987).

\bibitem{DPP} V.~Dotsenko, M.\ Picco and P.\ Pujol,
\newblock {\it Nucl.\ Phys.}~{\bf B455}, 701 (1995).

\bibitem{Rasmussen} J.~Rasmussen,
\newblock hep-th/0409026.

\bibitem{Ludwig2} E.~Bettelheim, I.~A.~Gruzberg,A.~W.~W.~Ludwig and P.~Wiegmann,
 \newblock {\it Phys. Rev. Lett.}~{\bf 95}, 251601 (2005).

\bibitem{Pasquier} V. Pasquier, {\it Nucl. Phys.}~{\bf B285}, 162 (1987).

\bibitem{DFSZ} P. Di Francesco, H. Saleur and J.-B. Zuber, {\it Nucl. Phys.}~{\bf B300}, 393 (1988).

\bibitem{Fendley2006} P.~Fendley, {\it J. Phys.}~{\bf A39}, 15445 (2006).

\bibitem{Cardy2008}  Y.~Ikhlef and J.~Cardy, arXiv:0810.5037.

\bibitem{Rajapour} M.~A.~Rajabpour, {\it J. Phys.}~{\bf A41}, 405001 (2008).

\bibitem{Raoul} R.~Santachiara, 
\newblock {\it Nucl. Phys.}~{\bf B793}, 396 (2008).

\bibitem{MarcoRaoul} M.~Picco and R.~Santachiara, 
\newblock {\it Phys. Rev. Lett.}~{\bf 100}, 015704 (2008).

\bibitem{MarcoRaoul2} M.~Picco and R.~Santachiara, 
\newblock in preparation.

\bibitem{Gamsa} A.~Gamsa and J.~Cardy, {\it J. Stat. Mech.} P08020 (2007).

\bibitem{Coniglio} A.~Coniglio, C.~Nappi, F.~Perrugi and L.~Russo, {\it J. Phys.}~{\bf A10},
205 (1977).

\bibitem{Sykes} M.~F.~Sykes and D.~S.~Gaunt {\it J. Phys.}~{\bf A9}, 2131 (1976).

\bibitem{Muller} H.~M\"uller-Krumbhaar, {\it Phys. Lett.}~{\bf 50A}, 27 (1974).

\bibitem{DHMMPW} Vl.~S.~Dotsenko, G.~Harris, E.~Marinari, E.~Martinec, M.~Picco and P.~Windey,
{\it Nucl. Phys.}~{\bf B448}, 577 (1995).

\bibitem{Gruzberg} I.~Rushkin, E.~Bettelheim, I.~A.~Gruzberg and P.~Wiegmann,
{\it J. Phys.}~{\bf A40}, 2165 (2007).

\bibitem{Zama_lat} A.~B.~Zamolodchikov, 
\newblock {\it Zh.~Eksp.~Teor.~Fiz.}~{\bf 75} (1978), 341 [{\it Sov.~Phys.~JETP}~{\bf 
48}, 168 (1978)].

\bibitem{Dotsi_lat} V.~S.~Dotsenko, 
\newblock  {\it Zh.~Eksp.~Teor.~Fiz.}~{\bf 75} (1978), 1083 [{\it Sov.~Phys.~JETP}~{\bf 
48}, 546 (1978)].

\bibitem{Potts_Clock} F.~Y.~Wu,
\newblock {\it Rev. of Mod. Physics}~{\bf 54}, 235 (1982).

\bibitem{Ashkin} J.~Ashkin and E.~Teller, 
\newblock {\it Phys. Rev.}~{\bf 64}, 178 (1943).

\bibitem{Lin} K.~.Y.~Lin and F.~Y.~Wu,
\newblock {\it J. Phys.}~{\bf C7}, L181 (1974).

\bibitem{Alcaraz2} F.~C.~Alcaraz and R.~K\"oberle,
\newblock {\it J.~Phys.}~{\bf A14}, 1169 (1981).

\bibitem{Zama_lat2} V.~A.~Fateev and A.~B.~Zamolodchikov, \newblock 
{\it Phys.~Lett.~JETP}~{\bf 92A}, 37 (1982).

\bibitem{Alcaraz} F.~C.~Alcaraz and R.~K\"oberle,
\newblock {\it J.~Phys.}~{\bf A13}, L153 (1980).

\bibitem{Alcaraz3} F.~C.~Alcaraz, 
\newblock {\it J.~Phys.}~{\bf A20}, 2511 (1987); {\it J.~Phys.}~{\bf A20},
623 (1987).

\bibitem{Rajabpour} M.~A.~Rajabpour and J.~ Cardy, \newblock arXiv:0708.3772

\bibitem{Wolff} U. Wolff, {\it Phys. Rev. Lett.}~{\bf 62}, 361 (1989).

\bibitem{FK} C. M. Fortuin and P. W. Kasteleyn, {\it Physica}~{\bf 57}, 536 (1972).

\bibitem{Stauffer} D. Stauffer and A. Aharony, Introduction to Percolation Theory, 2nd
edition (Taylor and Francis, London, 1994).

\end{thebibliography}
\end{document}